\documentclass[floatfix,prd,twocolumn,letterpaper,lengthcheck,superscriptaddress,showpacs,amssymb,amsmath,amsfonts,aps,nofootinbib,nopreprintnumbers,showpacs]{revtex4-1}
\usepackage{xcolor}
\usepackage[colorlinks, pdfborder={0 0 0}]{hyperref}
\usepackage{graphicx}
\usepackage[varg]{txfonts}
\usepackage{etoolbox}
\usepackage{makecell}
\usepackage{siunitx}
\apptocmd{\thebibliography}{\raggedright}{}{}
\definecolor{LinkColor}{rgb}{0.75, 0, 0}
\definecolor{CiteColor}{rgb}{0, 0.5, 0.5}
\definecolor{UrlColor}{rgb}{0, 0, 0.75}
\hypersetup{linkcolor=LinkColor}
\hypersetup{citecolor=CiteColor}
\hypersetup{urlcolor=UrlColor}
\def\area{\area{Probability of \\  consistency with \\area theorem}}
\newcommand{\gwparams}{\ensuremath{\vec{\vartheta}}}
\newcommand{\ip}[2]{\left< #1, #2 \right>}

\definecolor{maroon}{RGB}{128,0,0}

\begin{document}

\title[]{Model systematics in time domain tests of binary black hole evolution}

\author{Shilpa Kastha}
\affiliation{Albert-Einstein-Institut, Max-Planck-Institut f{\"u}r Gravitationsphysik, Callinstra{\ss}e 38, 30167 Hannover, Germany}
\affiliation{Leibniz Universit{\"a}t Hannover, 30167 Hannover, Germany}

\author{Collin D. Capano}
\affiliation{Albert-Einstein-Institut, Max-Planck-Institut f{\"u}r Gravitationsphysik, Callinstra{\ss}e 38, 30167 Hannover, Germany}
\affiliation{Leibniz Universit{\"a}t Hannover, 30167 Hannover, Germany}
\affiliation{Department of Mathematics, University of Massachusetts, Dartmouth, MA 02747, USA}

\author{Julian Westerweck}
\affiliation{Albert-Einstein-Institut, Max-Planck-Institut f{\"u}r Gravitationsphysik, Callinstra{\ss}e 38, 30167 Hannover, Germany}
\affiliation{Leibniz Universit{\"a}t Hannover, 30167 Hannover, Germany}

\author{Miriam Cabero}
\affiliation{Department of Physics and Astronomy, The University of British Columbia, Vancouver, BC V6T 1Z4, Canada}

\author{Badri Krishnan}
\affiliation{Albert-Einstein-Institut, Max-Planck-Institut f{\"u}r Gravitationsphysik, Callinstra{\ss}e 38, 30167 Hannover, Germany}
\affiliation{Leibniz Universit{\"a}t Hannover, 30167 Hannover, Germany}
\affiliation{Institute for Mathematics, Astrophysics and Particle Physics, Radboud University, Heyendaalseweg 135, 6525 AJ Nijmegen, The Netherlands}

\author{Alex B. Nielsen}
\affiliation{Department of Mathematics and Physics, University of Stavanger, NO-4036 Stavanger, Norway}

\date{\today}

\begin{abstract}
	We perform several consistency tests between different phases of binary black hole dynamics; the inspiral, the merger and the ringdown on the gravitational wave events GW150914 and GW170814. These tests are performed explicitly in the time domain, without any spectral leakage between the different phases. We compute posterior distributions on the mass and spin of the initial black holes and the final black hole. We also compute the initial areas of the two individual black holes and the final area from the parameters describing the remnant black hole. This facilitates a test of Hawking’s black hole area theorem. We use different waveform models to quantify systematic waveform uncertainties for the area increase law with the two events. We find that these errors may lead to overstating the confidence with which the area theorem is confirmed. For example, we find $>99\%$ agreement with the area theorem for GW150914 if a damped sinusoid consisting of a single mode is used at merger to estimate the final area. This is because this model overestimates the final mass. Including an overtone of the dominant mode decreases the confidence to $\sim94\%$; using a full merger-ringdown model further decreases the confidence to $\sim85-90\%$. We find that comparing the measured change in area to the expected change in area yields a more robust test, as it also captures over estimates in the change of area. We find good agreement with GR when applying this test to GW150914 and GW170814.
\end{abstract}

\maketitle

\section{Introduction}
\label{intro}

The first observation of gravitational waves (GWs) by the two detectors of the Laser Interferometer Gravitational-wave Observatory (LIGO), GW150914~\cite{Discovery},  was inferred to be produced by two merging black holes (BHs) with masses $\sim36M_\odot$ and $\sim29M_\odot$ at a luminosity distance of $\sim 400$ Mpc.
Following the first detection, several others~\cite{nitz20213ogc, GW151226, GW170104, GW170608, GW170814,GWTC1,Venumadhav:2019lyq,Zackay:2019btq, GWTC2, GWTC2population,GWTC22021search, 2021gwtc3} have also been detected during the three observing runs of the LIGO-VIRGO detectors to date. During the second observing run, GW170814 was detected coherently by the Advanced Virgo detector along with the two Advanced LIGO detectors. GW170814~\cite{GW170814} was produced by two BHs with masses $30^{+5.7}_{-3.0}M_{\odot}$~and~$25.3^{+2.8}_{-4.2}M_{\odot}$ at a luminosity distance of $540^{+130}_{-210}$Mpc.

These observations have enabled tests of the predictions of
Einstein's General Relativity (GR) in a highly nonlinear and
relativistic regime produced by compact binary mergers~\cite{GWTC-TGR,GWTC2-TGR}. 
Several tests of GR have been developed for
these recent GW detections, including parametrized tests of GR waveforms~\cite{AIQS06a,AIQS06b,Arun2012,YunesPretorius09,MAIS10,TIGER,  Li:2011cg,Meidam:2017dgf, Kastha_2019, Gupta_2020}, parametrized tests of the multipolar structure of binaries~\cite{Kastha2018, Kastha_2019}, tests of the parametrized post-Einsteinian formalism~\cite{YunesPretorius09,PPE2011}, inspiral-merger-ringdown (IMR) consistency tests~\cite{IMRConsistency}, and tests of black hole no-hair theorem~\cite{capano2021observation, Capano2020, Isi_2019, Dhanpal2019}.

The frequency-domain IMR consistency test is routinely performed on GW events detected by the LIGO and VIRGO detectors~\cite{GWTC-TGR}.
This test checks the consistency of the low frequency part of the observed signal with the high-frequency part. Results are presented in terms of the differences of the final mass and final spin inferred from the two different parts. These values are computed using extrapolations based on GR models parametrized in terms of the individual black hole masses and spins. Results to date are consistent with these differences being zero, implying that the GR based models are consistent with the data.

A related test uses measurements of the initial and final black holes' parameters as a check of the black hole area increase law.  This law states that within classical general relativity, assuming the null energy condition and cosmic censorship, the area of a black hole horizon can never decrease.  This was first shown by Hawking \cite{Hawking:1971tu, Hawking1972} and was later generalized to include non-differentiable event horizons and a cosmological constant \cite{Chrusciel:2000cu}.  Quasi-local versions of this law, which do not require cosmic censorship are also known \cite{Ashtekar:2002ag,Pook-Kolb:2019iao,Bousso:2015mqa}. 
Any stationary, astrophysical black hole is completely described by the Kerr metric in terms of its mass $M$ and angular momentum $J$ if the black hole no-hair theorem holds. The corresponding horizon area is $A=8\pi M^2 (1+\sqrt{1-\chi^2})$, where $\chi=J/M^2$ is the dimensionless spin of the black hole. For a binary black hole coalescence, the area theorem states that the final area $A_f$ of the merged black hole will be larger than the combined area ($A_1+A_2$) of the two initial black holes, 
\begin{equation}
\label{eq:a1a2}
A_1 + A_2 \equiv A_i < A_f.
\end{equation}
It is additionally expected that sufficiently removed from the dynamical regime, i.e. at very early or late times, these black holes will be well described by the stationary Kerr metric.

Following \cite{Hughes:2004vw}, a concrete proposal for a test of the area increase law was proposed in \cite{Cabero_2018}. The test consists of independent analyses of the early inspiral and the final ringdown stages, leading respectively to independent estimates of the initial and the final masses and spins of the binary. This test is carried out in the time domain, meaning that a portion of the strain data around the merger is excised in the time domain to separate the inspiral and ringdown phases. The data segment around the merger is excised in estimating the parameters, since a violation of the area theorem is perhaps most likely to occur near the merger of the two black holes where the spacetime is highly dynamical. These estimates are then used to obtain the areas of the individual BHs using the Kerr formula. To demonstrate the method, the authors used a simulated GW150914-like binary black hole signal and found that the area theorem could be confirmed at $75\%$ probability. 

Recently, Ref.~\cite{Isi:2020tac} has explored the validity of the area theorem in the time domain and presented observational evidence that actual GW150914 data is consistent with the theorem with a probability of 97\% when they do not excise the merger and 95\% when they excise 3ms of the merger. They also provide an estimate of the same by truncating the inspiral at different times before the peak amplitude and find that the different measurements support the
area theorem with probabilities within 88-97\%.

In this paper we apply the test proposed in \cite{Cabero_2018} to two GW events, GW150914 and GW170814. To test the area theorem or the IMR consistency between different phases of the waveform, we need each of these different phases (inspiral/ringdown) to have reasonable signal-to-noise ratio (SNR) to perform the parameter estimations. We choose GW150914 as it has a high network-SNR of $\sim 24$ \cite{nitz20213ogc, Discovery}, with a ringdown
SNR $\sim 8.5$~\cite{TOG} 3ms after the merger. 
The other event, GW170814, showed some support for a deviation from GR in initial analyses ($\Delta M_f/\bar{M}_f$ showed a second peak at higher values away from zero; see Fig. 2 of \cite{GWTC-TGR}).
To explore the validity of the area theorem we choose GW170814 as our second candidate event.

We separately analyze the data segment before a truncation time, removing later data, to obtain the initial parameters. Similarly, the data before a truncation time are removed, and the remainder is analyzed to estimate the final parameters of each event.
The truncation times may be chosen differently between the pre- and post-truncation analyses such that the merger phase is excluded.
We employ various waveform models for estimating the parameters to show the effect of waveform systematics on the constraints of the area increase law. We show that ignoring higher overtones in the ringdown waveform model leads to overestimation of the final mass. This yields an inconsistency between the estimated parameters from the pre- and post-truncation analyses, but perversely results in a \emph{better} agreement with the area theorem. A positive change in area, while obeying the area increase law, may still disagree with GR predictions if the area increase is too large. We use the ratio between the measured and the expected change of area, $\mathcal{R}=(A^{measured}_{f} - A^{measured}_i)/(A^{expected}_{f} - A^{measured}_{i})$, as a measure here. For a perfect measurement, obeying GR predictions, $\mathcal{R}=1$. However, in the presence of noise, we expect $\mathcal{R}$ to follow a Gaussian distribution with unit mean. In order to quantify the agreement of our results with GR, we also compute the mismatch, $C_{\mathcal{R}}$. 
It denotes the probability of $R$ lying within the range symmetric about the median of the $R$-distribution, extending to R=1. For GW150914, we find $C_{\mathcal{R}}=6.5\%$ when we avoid 26ms of data around the merger. Although, for GW170814, we obtain a better measurement yielding $C_{\mathcal{R}}=2.7\%$ when $\sim 13$ms of data is avoided around the merger.

This paper is organized as follows. In Sec.~\ref{sec2} we explain the method adopted here to excise the data and perform parameter estimation. In Sec.~\ref{sec3} and Sec.~\ref{sec4} we develop the different pre-truncation and post-truncation analyses. In Sec.~\ref{sec5} we discuss our results for the inspiral-merger-ringdown consistency test and in Sec.~\ref{sec6} for the test of the area theorem. Our concluding remarks are presented in Sec.~\ref{sec7}.

\section{Data preparation and parameter estimation} \label{sec2}

The gravitational wave strain, $h(t)$, observed by a detector can be schematically written as
\begin{align}
h(t-t_0) &= F_+(\alpha,\delta,\psi)h_+(t-t_0,\phi_0) \\
& + F_\times(\alpha, \delta, \psi)h_\times(t-t_0,\phi_0)\, ,
\end{align}
where $F_{+,\times}$ are the antenna pattern functions of the detector. The right ascension, $\alpha$, and declination, $\delta$, define the sky-location of the source in a geocentric coordinate system and the polarisation angle,
$\psi$, defines the relative orientation of the wave frame with respect to the
geocentric coordinate system~\cite{Apostolatos:1994mx,Jaranowski:1998qm}. $t_0$ is the arrival time of the signal at the detector and $\phi_0$ is the phase at $t_0$. $h_+(t)$ and $h_\times(t)$ are the two independent polarisations of the GW signal, given by
\begin{eqnarray}
h_+(t) &=& A_+(t)\cos\Phi(t) \,,\\
h_\times(t) &=& A_\times(t)\sin\Phi(t) \,,
\end{eqnarray}
where $A_{+,\times}$ are slowly varying amplitudes and $\Phi(t)$ is a
rapidly varying phase. 

\begin{figure*}[htp]
	\centering
	\begin{minipage}{\columnwidth}
		\centering
		\includegraphics[width=\textwidth]{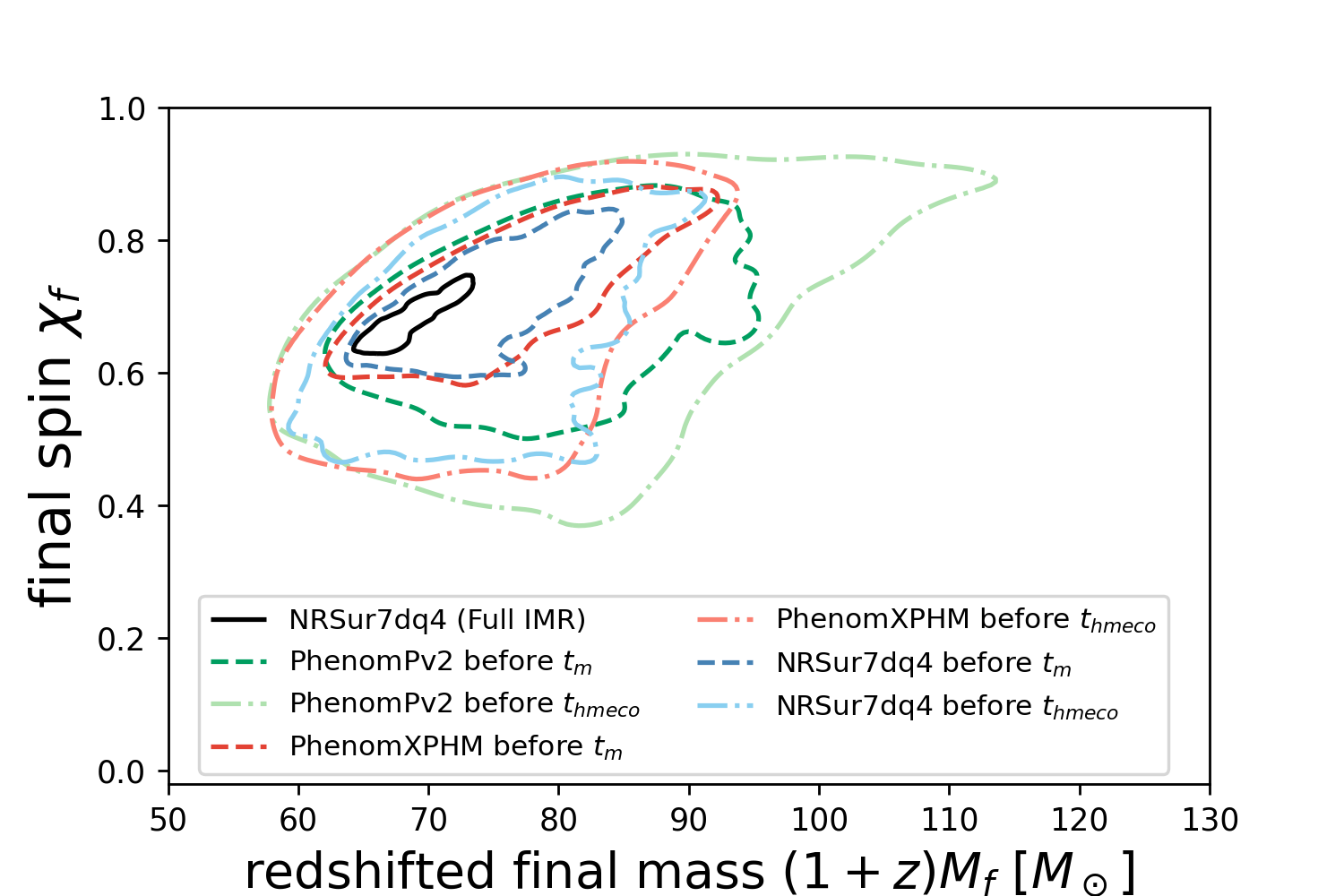}
	\end{minipage}%
	\begin{minipage}{\columnwidth}
		\centering
		\includegraphics[width=\textwidth]{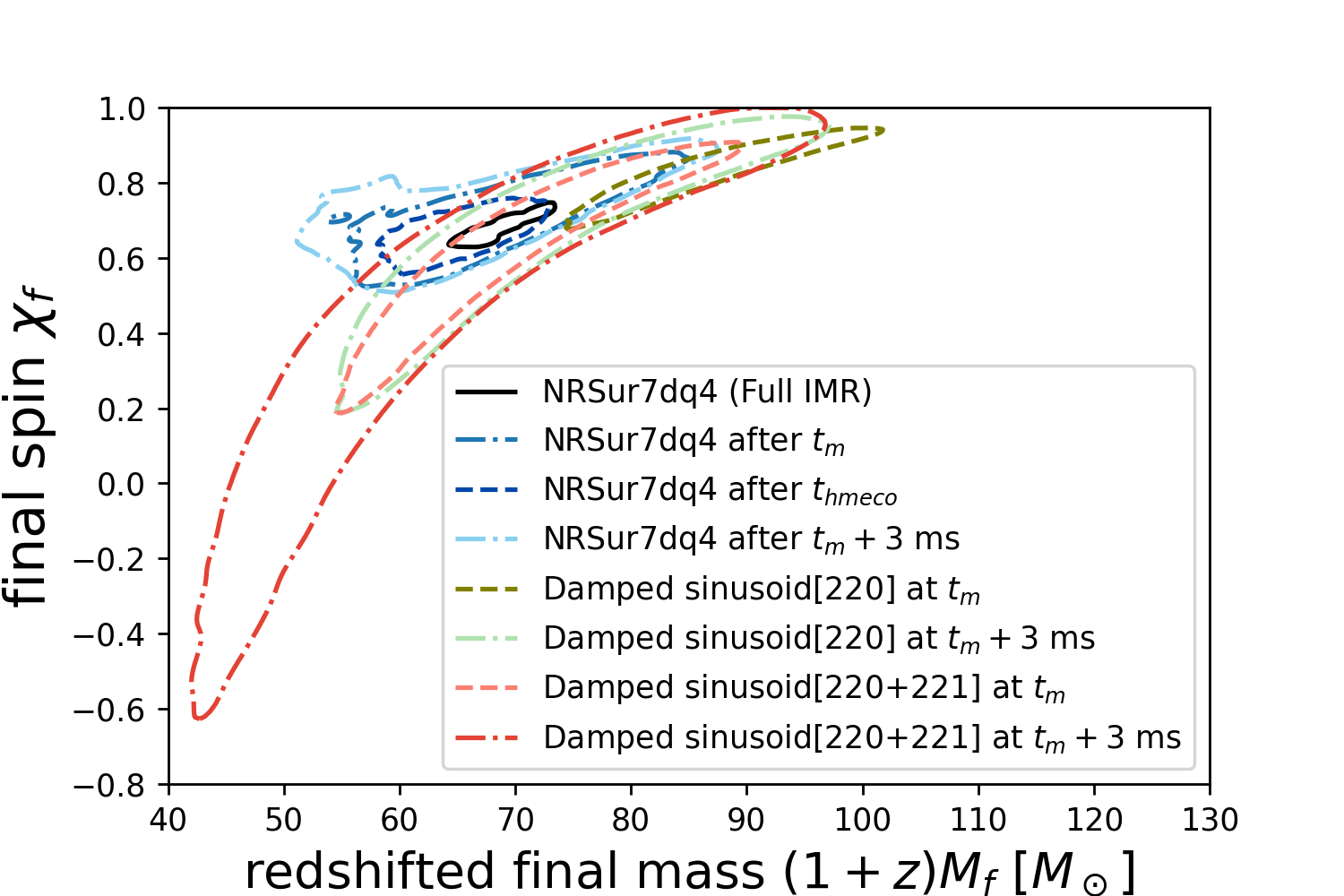}
	\end{minipage}
	\caption{The 90\% posterior contours of the redshifted final mass (detector frame) $M_f$ and the final spin $\chi_f$ obtained from the various pre- and post-truncation analyses for GW150914. In the left panel, we show the contours from the two pre-truncation analyses performed with two {\it gate-start-times}, $t_m$ and $t_{hmeco}$ using three different waveform models, IMRPhenomPv2, IMRPhenomXPHM and NRSur7dq4. On the right panel, we show the final mass-final spin contours obtained using damped sinusoids with only the dominant mode [220] and the dominant mode with one overtone [220+221] at different analysis start times (or the ``gate-end-times") as the representatives of the post-truncation analyses. We also show three post-truncation analyses with the NRSur7dq4 model for the three ``gate-end-times", $t_{hmeco}$, $t_m$, and $t_m+3$ms. The posterior contour arising from the full IMR analysis with NRSur7dq4 is shown by the innermost contour. The full IMR contour is contained within the contours of all the truncated analyses. }\label{Fig:GW150914_IMR}
\end{figure*}

To estimate the source parameters of a GW signal present in a given
data stream, $s(t)$, we use
Bayesian inference. We first consider a model for the signal $h$ within GR, parametrized by the source properties such as masses, spins etc., $\{M_1, \chi_1, ...\} \equiv \gwparams$
. According to Bayes' Theorem, the probability distribution of the model parameters given the data $s(t)$, $p(\gwparams|s,h)$ (known as
the \emph{posterior} distribution) is proportional to the likelihood, $\mathcal{L}(s|\gwparams,h)$, of observing the data given $\gwparams$ multiplied by a \emph{prior} distribution, $p(\gwparams)$, on the parameters representing the allowed range and expected distribution on \gwparams{}. For a network of $N$ gravitational-wave detectors $d$, the likelihood function is given by
\begin{equation}
\label{eqn:likelihood}
\mathcal{L}(s|\gwparams,h) \propto
\exp\left[-\frac{1}{2} \sum_{d=1}^{N}
\ip{s_d-h_d(\gwparams)}{s_d-h_d(\gwparams)}
\right],
\end{equation}
where $s_d$ is the data in the $d$-th detector (assuming the noise to be uncorrelated, stationary and Gaussian)  and $h_d$ is the waveform model (or
\emph{template}). 
Here, the noise-weighted inner product is defined as
\begin{equation}
\langle a| b\rangle =2 \,{\rm Re}\int_{f_{\rm low}}^{f_{\rm
		\rm high}}\frac{\tilde{a}(f)\,\tilde{b}^{*}(f)+\tilde{a}^{*}(f)\,\tilde{b}(f)}{S_n^{(d)}(f)}\,df,
\end{equation}
where $S_n^{(d)}(f)$ is the noise  power spectral density (PSD) of the $d$-th detector and $\tilde{a}(f)$ and $\tilde{b}(f)$ are frequency domain representations of the time-domain data. We use \texttt{PyCBC Inference}~\cite{Biwer_2019} to evaluate Eq. \eqref{eqn:likelihood} over the
large, multidimensional parameter space defining the waveform model. To sample the parameter space we use the $\texttt{dynesty}$~\cite{speagle:2019} and parallel-tempered $\texttt{emcee}$~\cite{ForemanMackey:2012ig,Vousden:2015} stochastic samplers. Marginalizing the resulting distribution yields measurements on individual parameters.

Our aim is to analyze separately the early and late parts of the signal
to investigate the consistency between the different phases, and to explore the validity of the area theorem. In order to perform these analyses, we excise portions of the templates, keeping a desired time segment intact to perform the parameter estimation.
As observed in ref.~\cite{Cabero_2018}, merely excising times from the template can lead to biases unless a corresponding part of the signal is also removed from the data. Hence, both the template and the signal need to be excised while performing parameter estimation. We accomplish this by doing ``gating and in-painting''~\cite{Zackay:2019btq,capano2021observation}. In this method, we zero-out (``gate'') the residual $s_d-h_d(\vec{\vartheta})$ over times to be excised, then add a component to the gated times such that the contribution of these times to the likelihood is zero (``in-painting''). This removes biases that arise due to the convolution of the inverse covariance of the detector noise with the residual. The end result is the pre-truncation analysis is independent of the post-truncation analysis. 

Since the sky location affects the
arrival time of the signal in the detectors, the specific time for the gating changes with respect to different sky locations. If the sky location is varied during the analysis, the truncated template favors those sky locations that include more of the signal. This results in an estimation of the parameters shifted from their true values~\cite{Cabero_2018}. To avoid this, here we consider a fixed sky location. In order to fix the values for $\alpha$ and $\delta$, we first perform the parameter estimation analysis on the full data stream, using the complete inspiral-merger-ringdown signal. We then use the corresponding maximum likelihood values for the sky location in the area-theorem analysis. Using slightly different values within the range of the posterior distributions on $\alpha$ and $\delta$
does not affect our final conclusions.

\section{Pre-truncation analysis}
\label{sec3}

In this section we focus on measuring the initial parameters from the early part of the data. To model the gravitational-wave signal in the early to late inspiral regime, we use three waveform models: IMRPhenomPv2~\cite{Husa_2016,Khan_2016}, which models precessing binaries; IMRPhenomXPHM~\cite{Pratten_2021, lalsuite}, which models precessing binaries with higher modes; and NRSur7dq4~\cite{Varma_2019}, which also models precessing binaries with higher modes, mass ratios $q\leq 6$, and spin magnitudes $\chi_1, \chi_2 \leq 0.8$.

\begin{figure*}
	\centering
	\begin{minipage}{\columnwidth}
		\centering
		\includegraphics[width=\textwidth]{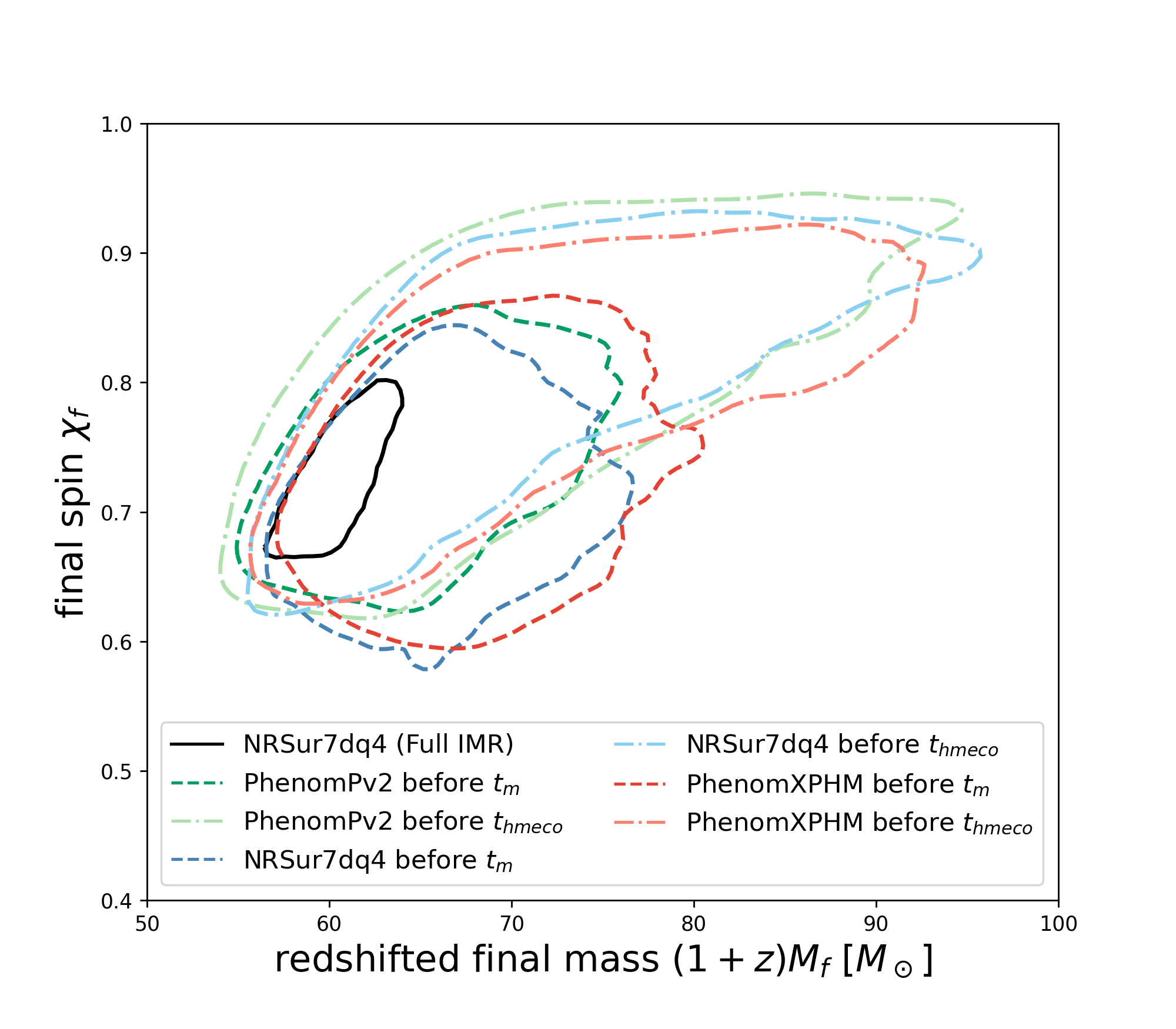}
	\end{minipage}%
	\hspace{0.3cm}
	\begin{minipage}{\columnwidth}
		\centering
		\includegraphics[width=\textwidth]{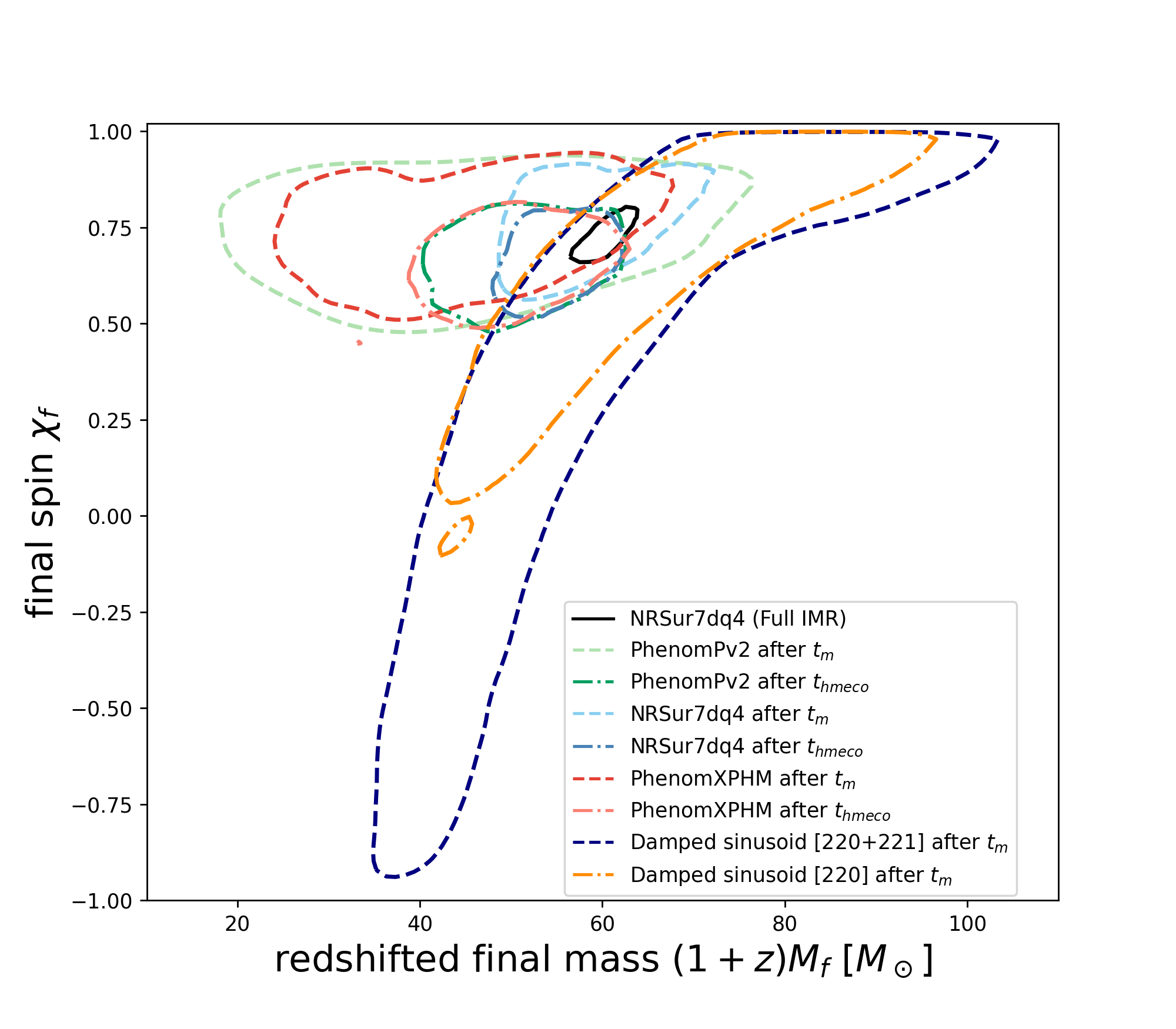}
	\end{minipage}
	\caption{The 90\% posterior contours of the redshifted final mass (detector frame) $M_f$ and the final spin $\chi_f$ obtained from the various pre- and post-truncation analyses for GW170814. On the left panel we show the contours from the pre-truncation analyses performed with two {\it gate-start-times}, $t_m$ and $t_{hmeco}$ using three different waveform models, IMRPhenomPv2, IMRPhenomXPHM and NRSur7dq4. On the right panel we show the 90\% posterior contours of the redshifted final mass (detector frame) $M_f$ and the final spin $\chi_f$ obtained using the damped sinusoids with only dominant mode [220] and the dominant mode with one overtone [220+221] at the merger time $t_m$ as the representatives of the post-truncation analyses for GW170814. We also show the results for the two post-truncation analyses with two ``gate-end-times", $t_m$ and $t_{hmeco}$ for IMRPhenomPv2, IMRPhenomXPHM and NRSur7dq4 waveform models. The ringdown contour is larger when the 221 mode is included because inclusion of the overtone increases the dimension of the parameter space. The posterior contour arising from the full IMR analysis with NRSur7dq4 waveform model is shown by the innermost solid black contour.
	}
	\label{Fig:GW170814_2}
\end{figure*}

For the pre-truncation analysis, we excise the residual after a desired time (``{\it gate-start-time}'') and perform the sampling on the remaining data segment. 
While the truncation can be started at any time within the inspiral regime, earlier truncation times lead to posterior distributions yielding worse constraints on the estimated parameters due to decreasing SNR of the remaining signal.
We choose two different {\it gate-start-times} in our analysis. In the first case, we use the merger time, $t_m$, defining the IMR waveform as the truncation time.
For GW150914, we base the merger time on the estimate in \cite{GWTC2-TGR}, while for GW170814 we use the results from \cite{nitz20213ogc}, as explained in more detail in the respective sections.

In the second case we use the time corresponding to the hybrid minimum energy circular orbit (hybrid MECO), $t_{hmeco}$~\cite{Cabero:2016ayq}. The hybrid MECO depends on the mass ratio and the spins of the black holes. It corresponds to a time earlier than $t_m$, which can be considered as the end of the inspiral phase for comparable mass binaries, but before the peak amplitude GW emission. Here the hybrid MECO time is computed using the maximum likelihood values for the masses and the spins of the individual black holes from the full IMR analysis.

In the parameter estimation performed here, we use uniform priors on merger time and the source-frame component masses. For the different analyses we use different prior ranges on the component masses. We keep the interval between the minimum and the maximum values of the component masses sufficiently large so that the posteriors have negligible values on the prior boundaries.

We also assume a distance prior uniform in comoving volume assuming a flat $\Lambda$CDM cosmological model. For the spins, we use uniform priors for the magnitude of the spin and isotropic for the orientation. We numerically marginalize over polarization.
From each of the inspiral analyses we obtain the posterior distributions on the individual masses and spins. Finally, using the fitting formula in Refs.~\cite{Healy_2014, Hofmann_2016, Jim_nez_Forteza_2017}, we convert the initial masses and spins to the final mass and spin posterior.

\section{Post-truncation analysis}
\label{sec4}

After the two individual black holes merge, the remnant is expected to settle down to a final stable black hole during the ringdown phase. 
Similar to the pre-truncation analysis, we perform post-truncation analyses using the GW signal emitted during this merger and post-merger phase. As opposed to the previous analysis, here we excise the residual all the way up to the analysis start-time (``{\it gate-end-time}") to perform sampling on the data segment after the {\it gate-end-time} to obtain the posterior distribution on the final parameters. 

We preform two sets of post-truncation analyses. In one, we use the late part of the same IMR waveform models as in the pre-truncation analysis to obtain the posterior distribution on the masses and the spins. Using the fitting formula in Refs.~\cite{Healy_2014, Hofmann_2016, Jim_nez_Forteza_2017} we then convert the component-object posteriors to a distribution on the final mass and spin of the remnant BH. For these cases we use similar priors as in the pre-truncation analyses: uniform priors on merger time and the source-frame component masses, and a distance prior uniform in comoving volume. For the spins, the priors are uniform in magnitude of the spin and isotropic for the orientation. We also numerically marginalize over polarization.

\begin{figure*}[htp]
	\centering
	\begin{minipage}{\columnwidth}
		\centering
		\includegraphics[width=\textwidth]{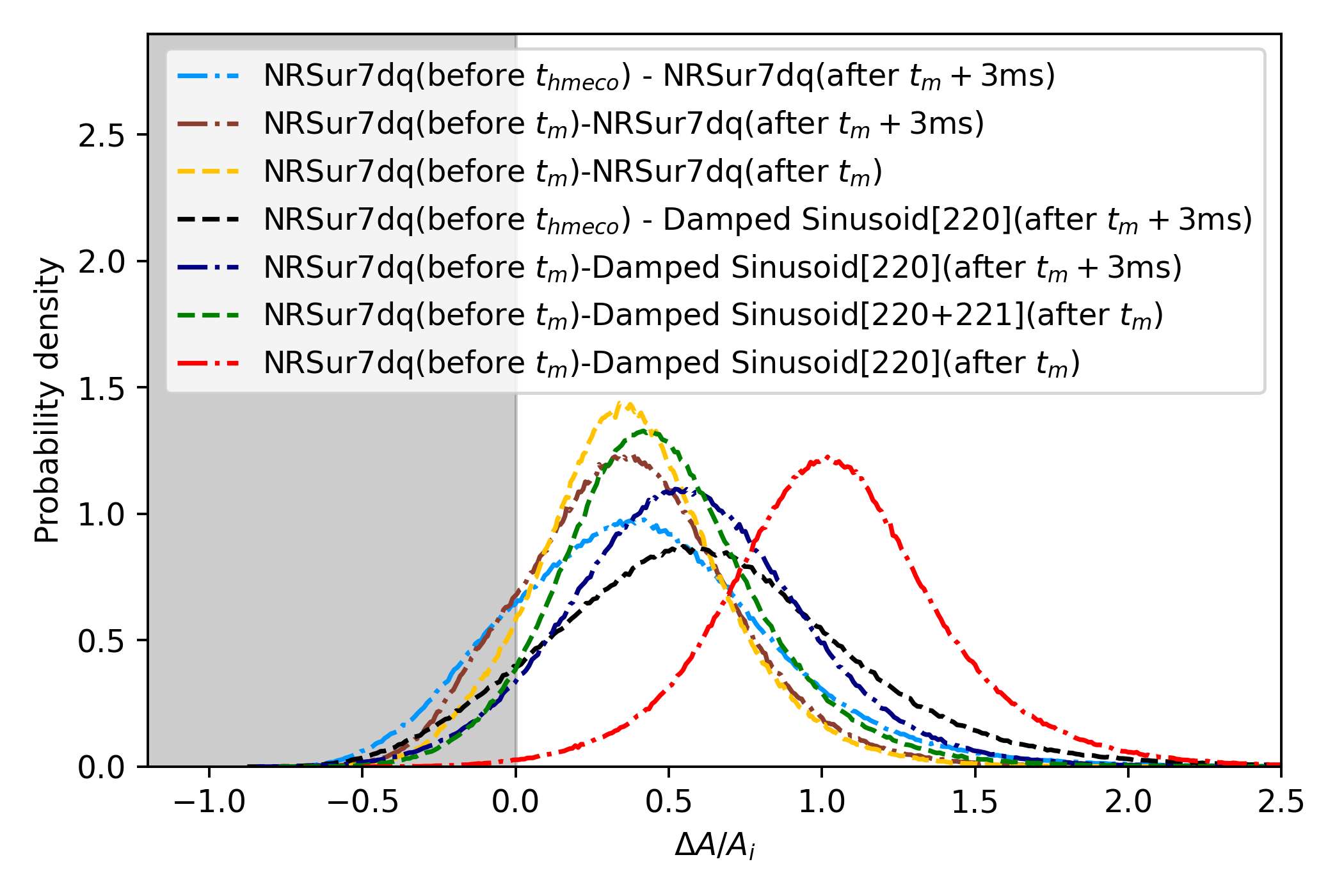}
		\label{Fig:GW150914_relative_area_change}
	\end{minipage}%
	\hspace{0.3cm}
	\begin{minipage}{\columnwidth}
		\centering
		\includegraphics[width=\textwidth]{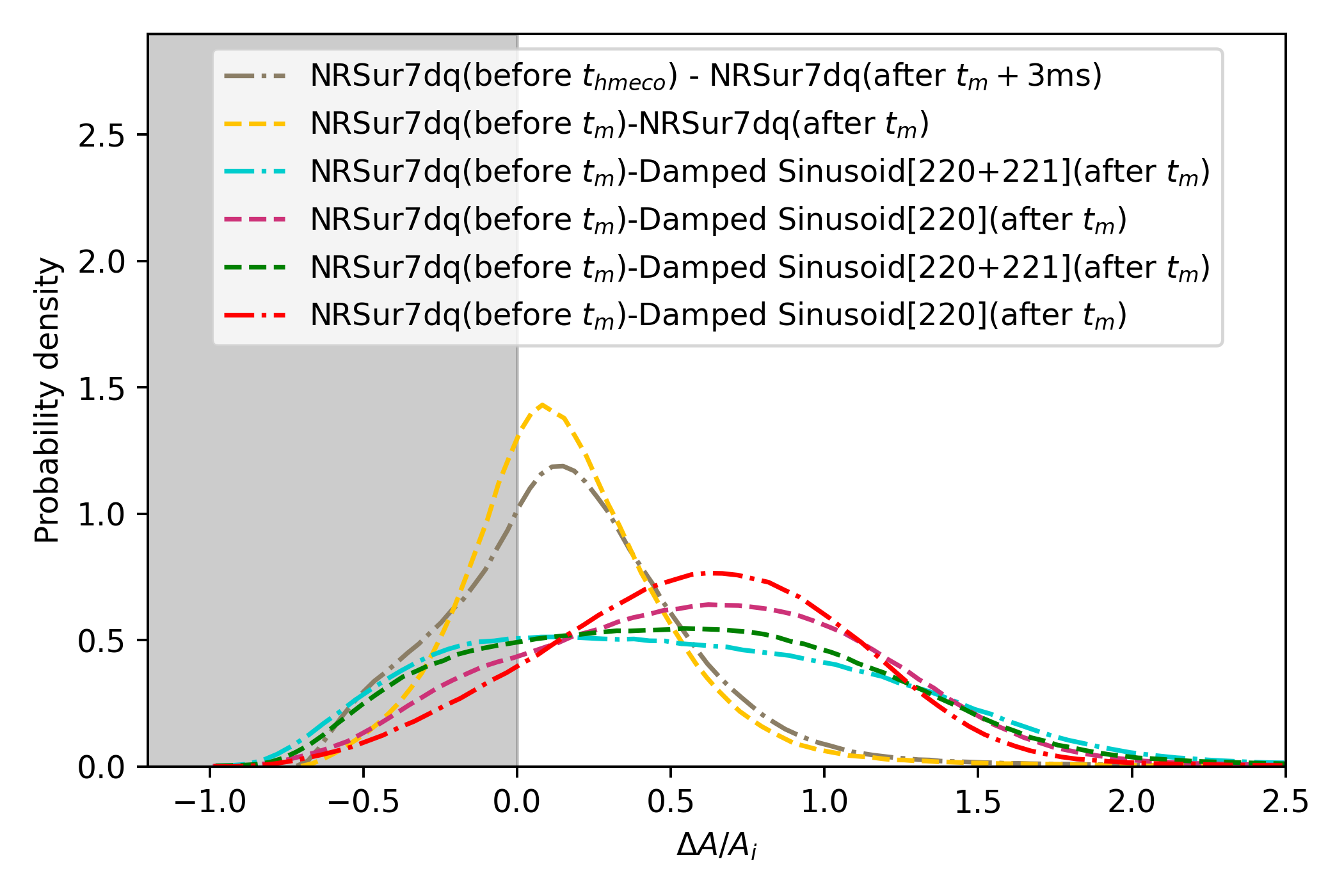}
		\label{Fig:GW170814_relative_area_change}
	\end{minipage}
	\caption{Fractional change in the black hole horizon area, $\Delta A/A^{measured}_i$, due to the BBH merger [GW150914 on the left and GW170814 on the right] for different pre- and post truncation analyses performed with NRSur7dq4 or the damped sinusoid models. The change in the area is denoted by $\Delta A=(A^{measured}_f-A^{measured}_i)$ with $A^{measured}_i$ being the initial area inferred using the parameters from the pre-truncation analysis and $A^{measured}_f$ is the final area measured with the estimated parameters form the post-truncation analysis.}
	\label{Fig:relative_area_change}
\end{figure*}

In the second set of post-truncation analyses we use damped sinusoids as the signal model. The signal emitted during this post-merger phase is conventionally called the ``ringdown'' signal and can be decomposed into a sum of exponentially damped sinusoids~\cite{Berti_2009}. The gravitational waveform for the ringdown can be schematically written in terms of spin-weighted spheroidal harmonics as 
\begin{equation} 
h_+ + i h_\times = \sum_{\ell,m,n} {}_{-2}S_{\ell m n}(\iota,\varphi) A_{\ell mn}
e^{i (\Omega_{\ell mn}t + \phi_{\ell m n})}\,,
\end{equation} 
where the sum is over the various integer quantum numbers, $l \ge 2$, $-\ell \le m \le \ell$ and $n \ge 0$ denoting the different quasi-normal modes. $\iota$ is the inclination angle, $\varphi$ is the azimuthal angle of the black hole with respect
to the observer, and ${}_{-2}S_{\ell m n}(\iota,\varphi)$
are the spin-weighted spheroidal harmonics.
The spin-weighted spheroidal harmonics reduce to the usual spin-weighted spherical harmonics, $Y_{\ell m}$ for the non-spinning case.
The various mode amplitudes $A_{\ell mn}$ and the phases $\phi_{\ell m n}$
depend on the initial configuration of the binary and on the particular theory of gravitation. In our analysis we use spheroidal harmonics~\cite{capano2021observation} and treat the mode amplitudes and the phases as independent unknown parameters. The complex frequencies $\Omega_{\ell m n}$ consist of the quasi-normal mode frequencies ($f_{\ell mn}$) and the damping times ($\tau_{\ell mn}$), which can be determined from the Teukolsky equation~\cite{Teukolsky:1972my,Leaver:1985ax}. The no-hair theorem states that all ($f_{\ell mn}$) and ($\tau_{\ell mn}$) are determined by only two quantities, the mass $M_f$ and spin $\chi_f$ of the black hole.
Here, we use $\chi_f \in (-0.99, 0.99)$, with positive or negative values referring to perturbations that are co- or counter-rotating with respect to the black hole's spin.

One can in principle vary many modes in the ringdown analysis with Bayesian inference. However, this increases the dimension of the parameter space leading to weaker constraints on the measured parameters if the contribution to the signal from additional modes is weak. Here we perform two types of ringdown analyses for each event. First we consider only the dominant fundamental mode, for which $\ell=m=2$ and $n=0$ (i.e., the [220] mode). Secondly, we include one overtone of the dominant mode (i.e., [220+221]), giving a signal template consisting of two modes. In a similar study of GW150914 in Ref.~\cite{Isi_2019}, the authors claimed the existence of the fundamental quasinormal mode and one overtone associated with the dominant angular mode ($\ell= m = 2$) with 3.6$\sigma$ confidence.

In the ringdown analyses performed here, we vary the following parameters: final mass $M_f$, final spin $\chi_f$,  $A_{2 2 0}$, $\phi_{220}$, and inclination $\iota$ in the single-mode 220 case, and add two additional parameters $A_{2 2 1}$ and $\phi_{221}$ when considering the additional first overtone. We numerically marginalize over the polarization $\psi$. Furthermore, we use different analysis start times (or ``{\it gate-end-times"}) for the ringdown template. For each of these cases, the priors on final mass and final spin are assumed to be uniform in the following ranges 
\begin{align} 
&M_{f} \in [10, 200)\\
&\chi_{f} \in [-0.99, +0.99)\,.
\end{align}
For $A_{2 2 0}$, we choose a prior uniform in $\log_{10}$.
We allow the overtone amplitude, $A_{2 2 1}$, to uniformly vary from zero to ten times that of $A_{2 2 0}$ when we start the analysis at the merger time, $t_m$.
When starting the analysis at later times, we choose a uniform prior so that $A_{2 2 1}<A_{2 2 0}$.

\section{Consistency between Inspiral-merger-ringdown phases}
\label{sec5}
Our results for GW150914 are summarized in Fig.~\ref{Fig:GW150914_IMR}. In the left panel of Fig.~\ref{Fig:GW150914_IMR}, we compare the posterior distributions from the various pre-truncation analyses performed with different segments of the data and different waveform models. The innermost contour refers to the full IMR analysis performed using the complete data segment with the NRSur7dq4 model. We find similar estimates with the IMRPhenomXPHM and IMRPhenomPv2 models. For all these cases we fix the sky location at $\alpha=1.252$, $\delta=-1.224$, the maximum-likelihood values from \cite{Nitz:2020oeq}.

For the pre-truncation analyses we consider two different gate start times, after which the residual is excised for each of the three waveform models. The first one is the merger time $t_m$, where we use the estimate from \cite{GWTC2-TGR}, $\SI{1126259462.423 }{\second}$ GPS time at the LIGO Hanford site.
Using the sky-location quoted above, this corresponds to $\SI{1126259462.411}{\second}$ in geocentric GPS time.
The second one is the hybrid MECO time, $t_{hmeco}=\SI{1126259462.388}{\second}$ in geocentric GPS time, which is $\sim$23ms earlier than $t_m$. Different waveform models provide different constraints on both the final mass and final spin parameters, but all of them are consistent with the full IMR analysis.

\begin{table*}[htp]
\begin{center}
\begin{tabular}{|| l | c | c | c ||}
 \hline
 \textbf{GW150914} &\shortstack{ $\Delta A/A^{measured}_i $\\ (NRSur7dq4)} &\shortstack{$\Delta A/A^{measured}_i $\\ (IMRPhenomXPHM)} & \shortstack{$\Delta A/A^{measured}_i $\\ (IMRPhenomPv2)} \\ [0.5ex] 
 \hline\hline
 IMR model before $t_m$-Damped Sinusoid [220] after $t_m$ & $1.03^{+0.66}_{-0.58}(99.7\%)$ & $0.99^{+0.69}_{-0.66} (99.2\%) $& $0.83^{0.74}_{-0.82}$(95.1\%) \\
 \hline
 IMR model before $t_m$-Damped Sinusoid [220+221] after $t_m$ & $0.46^{+0.61}_{-0.48}$(94.3\%)
 & $0.43^{+0.63}_{-0.52} (91.3\%)$ & $0.31^{+0.66}_{-0.60}$(79.3\%) \\ 
 \hline
 IMR model before $t_m$-Damped Sinusoid[220]  after $t_m+$3 ms & $0.56^{+0.66}_{-0.60}$(93.8\%)
 & $0.52^{+0.68}_{-0.62}$(91.4\%) & $0.38^{+0.73}_{-0.67}$(81.3\%)  \\
 \hline
 IMR model before $t_m$-Damped Sinusoid [220+221] after $t_m+$3 ms & $0.29^{+0.81}_{-0.64}$(74.6\%)& $0.25^{0.83}_{-0.64}(71.7\%)$& $0.13^{+0.85}_{-0.62}$(61.0\%)\\
 \hline
 IMR model before $t_m$-IMR model after $t_m$ & $0.37^{+0.54}_{-0.47}$(90.5\%) & $0.33^{+0.59}_{-0.51}(85.5\%)$ &$0.20^{+0.61}_{-0.57}(71.7\%)$ \\ 
 \hline
 IMR model before $t_m$-IMR model after $t_m+$3 ms & $0.37^{+0.59}_{-0.52}$(86.6\%) & $0.31^{+0.66}_{-0.58}$(80.0\%) & $0.18^{+0.68}_{-0.59}(67.6\%)$ \\ 
 \hline
 IMR model before $t_{hmeco}$-Damped Sinusoid [220] after $t_m$ &$1.1^{+0.94}_{-0.85}$(98.6\%) & $1.08^{+0.98}_{-0.99}$(96.3\%)& $0.76^{+1.12}_{-1.01}$(85.2\%) \\
 \hline
 IMR model before $t_{hmeco}$-Damped Sinusoid [220+221] after $t_m$ & $0.51^{+0.81}_{-0.64}$(90.1\%) & $0.49^{+0.83}_{-0.73}(87.2\%)$&$0.25^{+0.92}_{-0.73}$(67.7\%) \\
 \hline
 IMR model before $t_{hmeco}$-Damped Sinusoid[220] after $t_m+$3 ms & $0.6^{+0.87}_{-0.73}$(90.6\%) & $0.58^{+0.90}_{-0.81}$(87.7\%) & $0.32^{+1.01}_{-0.77}(70.0\%)$ \\ 
 \hline
 IMR model before $t_{hmeco}$-Damped Sinusoid [220+221] after $t_m+3$ ms & $0.4^{+0.78}_{-0.63}$(83.7\%) & $0.29^{+1.02}_{-0.74}$(71.4\%) & $0.06^{+1.09}_{-0.66}$(54.4\%) \\
 \hline
 IMR model before $t_{hmeco}$--IMR model after $t_m$ & $0.41^{0.73}_{-0.61}$(86.4\%) & $0.38^{+0.78}_{-0.68}$(82.3\%) & $0.16^{+0.86}_{-0.67}$(61.7\%) \\
 \hline
 IMR model before $t_{hmeco}$-IMR model after $t_m+$3 ms & $0.4^{+0.78}_{-0.63}$(83.6\%) & $0.36^{+0.85}_{-0.72}$(78.0\%) & $0.12^{+0.93}_{-0.67}(59.0\%)$ \\ 
 \hline
 & & &\\[0.9ex] 
 \hline\hline
 \textbf{GW170814} & &  & \\ [0.5ex] 
 \hline\hline
 IMR model before $t_m$-Damped Sinusoid [220+221] after $t_m$ & $0.52^{+1.12}_{-0.95}$(77.3\%)&   $0.45^{+1.1}_{-0.90}$(74.9\%) & $0.67^{1.21}_{-1.05}(81.3\%)$\\ 
 \hline
 IMR model before $t_m$-Damped Sinusoid [220] after $t_m$ & $0.63^{+0.81}_{-0.87}$(87.8\%)& $0.55^{+0.81}_{-0.83}(85.5\%)$& $0.8^{+0.86}_{-0.99}$(90.4\%)\\ 
 \hline
 IMR model before $t_m$-IMR model after $t_m$ & $0.16^{+0.66}_{-0.46}$(71.8\%) & $-0.33_{-0.41}^{+0.67}$(17.4\%) & $-0.23^{1.15}_{-0.47}$(31.9\%)\\ 
 \hline
 IMR model before $t_{hmeco}$-IMR model after $t_m$ & $0.17^{+0.71}_{-0.59}(69.1\%)$ & $-0.35_{-0.42}^{+0.75}$(19.1\%) & $-0.31^{1.17}_{-0.45}$(27.6\%)\\ 
 \hline
 IMR model before $t_{hmeco}$-Damped Sinusoid [220+221] after $t_m$ & $0.49^{+1.26}_{-0.96}(74.7\%)$ &  $0.40^{+1.29}_{-0.93}$(70.8\%) & $0.52^{1.29}_{-1.01}(74.7\%)$\\ 
 \hline
 IMR model before $t_{hmeco}$-Damped Sinusoid [220] after $t_m$ & $0.62^{+0.93}_{-0.94}(83.9\%)$ & $0.53^{+0.97}_{-0.94}$(79.6\%) & $0.66^{+0.94}_{-1.02}(83.5\%)$\\ 
 \hline
\end{tabular}
\caption{The median values and 90\% bounds on the fractional change of the BH horizon areas with different IMR waveform models at different truncation times. The first column corresponds to the different combinations of pre- and post-truncation analyses as indicated by the various rows. Here ``IMR model" indicates the different IMR waveform models used in parameter estimation, which is denoted in the header of the remaining three different columns. The percentages of the posteriors that have positive area increase are given in parentheses. For GW170814, due to low SNR ($\sim 7$) of the signal in post-truncation segment after $t_m$, we find larger posterior bounds on the parameters from IMRPhenomXPHM and IMRPhenomPv2 model, resulting in negative relative change in black hole horizon area.}\label{Area-table} 
\end{center}
\end{table*}

For a fixed waveform model, we obtain better constraints for the ``gate-start-time" at $t_m$ (denoted by ``IMR model before $t_m$'') compared to $t_{hmeco}$ (``IMR model before $t_{hmeco}$'') in left panel of Fig.~\ref{Fig:GW150914_IMR}). This is because later times include signal power from the late inspiral regimes. The posterior contours shrink and converge to the full IMR values as more data are included. Comparing between different waveform models, we find that the strongest constraints are obtained with the NRSur7dq4 model. 

The different post-truncation analyses are shown in the right panel of Fig.~\ref{Fig:GW150914_IMR}. First, we consider the same waveform models as in the pre-truncation analyses, where we excise the residual before $t_{hmeco}$, $t_m$ and $t_m+3\text{ms}$. We find that all three waveform models provide consistent results for each of the gate-end-times. Hence we only show the posterior arising from the NRSur7dq4 model. As is evident from the figure, the posterior distribution arising from choosing $t_{hmeco}$ as ``gate-end-time" provides the most stringent constraints on the parameters and the closest to the IMR values among these three. This is expected since in this case we exclude only the early inspiral segment of the data and include the late inspiral, merger and the ringdown phases.

Secondly, we perform the ringdown analysis as described in Sec.~\ref{sec4}. Here we choose two different start times for the ringdown models, $t_m$ and $t_m+$3 ms and for each of these, we do a single mode [220] analysis with only the dominant $\ell=m=2, n=0$ harmonic and another with the dominant harmonic and the first overtone [220+221], $\ell=m=2, n=1$. We find that at $t_m$, the dominant mode [220] analysis provides estimates with higher final mass and higher final spin than the IMR result. The 90\% posterior contours are completely disjoint. This is expected: applying the ringdown analysis too early leads to too low a ringdown frequency and results in overestimating the final mass. However, including the first overtone we recover estimates consistent with the full IMR analysis. Starting both the single- and two-mode analyses at 3ms, we find that the dominant mode analysis provides tighter constraints compared to the [220+221] analysis.

We find no support in our posterior for final masses above $100 M_\odot$, but masses below $50 M_\odot$ are supported for the quasi-normal mode analysis after 3ms with [220+221]. We also find that as we end the truncation at later times, the remaining signal becomes quieter and the constraints on final mass and final spin get broader.
In comparison with Isi et al. \cite{Isi:2020tac}, we use the same merger time, $t_m = \SI{1126259462.423}{\second}$ GPS time at the Hanford detector site, but perform the analysis with a different sky location (see Ref.~\cite{Isi_2019} for comparison). We find that changing the sky location does not affect our final result.
We also find similar final mass and final spin estimates from the [220]-analysis at $t_m$ and the [220+221] analysis at $t_m + \SI{3}{\milli\second}$ as quoted in~\cite{Isi:2020tac}. However, the estimates from the pre-truncation analysis carried out before $t_m$ with NRSur7dq4 differ from the results quoted in \cite{Isi:2020tac}.
In this particular case, we find the final mass and final spin estimates to be $(1+z)M_f\sim73.3^{+9.3}_{-6.6}M_\odot$ and $\chi_f\sim 0.72^{+0.08}_{-0.1}$ respectively.

We perform a similar set of analyses with GW170814 and present our results in Fig.~\ref{Fig:GW170814_2}. We fix the sky-location parameters of GW170814 to $\alpha=0.8$, $\delta=-0.8$, the maximum likelihood values from the IMR analysis in~\cite{nitz20213ogc}.
We use two truncation times to start or end the gating here, the merger time $t_m$ and hybrid MECO time $t_{hmeco}$.
For the merger time, we use the maximum likelihood value of the coalescence time $t_c$ from~\cite{nitz20213ogc}.
The hybrid MECO time is calculated from the waveform with the maximum likelihood parameters also from~\cite{nitz20213ogc}.
These are $t_m=\SI{1186741861.527}{\second}$ and $t_{hmeco}=\SI{1186741861.5136}{\second}$ in geocentric GPS time, corresponding to $t_m = \SI{1186741861.531}{\second}$ and $t_{hmeco}=\SI{1186741861.513}{\second}$ at the Hanford detector site.

The estimated final mass-final spin contours from the pre-truncation analyses before each of the gate-times are presented in the left panel of Fig.~\ref{Fig:GW170814_2}. We also show the IMR results from the analysis of the full data segment with NRSur7dq4 waveform model. This is found to be consistent with all the different analyses.

In the case of the pre-truncation analyses, for a fixed gate-start-time, the different waveform models provide very similar results. The post-truncation analyses are presented in the right panel of Fig.~\ref{Fig:GW170814_2}. Here we find that at fixed gate-end-time, the NRSur7dq4 model provides the most stringent constraints. In addition to these IMR waveform models we show our estimates from the two ringdown analyses with the [220] and the [220+221] modes performed at $t_m$ in right panel of Fig.~\ref{Fig:GW170814_2}. We find that both estimates are consistent with the IMR values. However the analysis with only the dominant mode provides better constraints on the parameters. Due to the post-merger signal being quiet, we do not find reasonable constraints on the final mass and final spin for any analysis performed beyond $t_m$ and omit them in the figure. 

\section{Area theorem}
\label{sec6}
In this section we investigate the validity of the area theorem from the estimated parameters described in the previous sections. In Table~\ref{Area-table} we report the 90\% bounds on the fractional change of the black hole horizon area, $\Delta A=(A^{measured}_f-A^{measured}_i)/A^{measured}_i$, for the five (four) different gate-times and three different waveform models for GW150914 (GW170814).  Here $A^{measured}_i=A^{measured}_1+A^{measured}_2$ is computed using the initial component masses and spins from the pre-truncation analyses and $A_f$ is the final area estimated from the post-truncation parameters.  
A graphical representation of the same is provided in Fig.~\ref{Fig:relative_area_change} for GW150914 (left panel) and GW170814 (right panel). In these figures we only provide the results obtained using combinations of NRSur7dq4 and the quasi-normal mode model for different data segments.

For GW150914 we obtain a 99\% agreement with the area theorem when NRSur7dq4 waveform model before $t_m$ is the pre-truncation model and a single [220] damped sinusoid after $t_m$ as the post-truncation model. The agreement reduces to 98\% when the pre-truncation analysis is limited to times before $t_{hmeco}$. This high agreement with the area theorem is due to the fact that the single [220] damped sinusoid model after $t_m$ overestimates the final mass and final spin, and hence overestimates the final area.

A weaker constraint $\sim 94\%$ on the validity of the area theorem is obtained when the pre-truncation analysis is carried out before $t_m$ with the post-truncation analysis being the simple ringdown analyses with [220+221] at $t_m$ and just [220] at $t_m+3$ms (see the second and the third entry in Table~\ref{Area-table}).
These two estimates are slightly lower than the results in Isi et al~\cite{Isi:2020tac}. Here, in the pre-truncation analysis before $t_m$, the prior ranges on the component masses (uniform between $11M_{\odot}-120M_{\odot}$) are chosen to ensure that the posteriors have negligible support at the prior boundaries. Using tighter priors (uniform between $17M_{\odot}-76M_{\odot}$, which excludes some part of posterior on the component masses), we find improved constraints on the area theorem, up to $\sim$97\%, as found in Ref.~\cite{Isi:2020tac}.
We also find that using uniform priors on the component masses provides similar results if we use uniform priors on total mass and mass ratio, as was done in Ref.~\cite{Isi:2020tac}. To be precise, we find that the difference in the prior distribution function (whether uniform on component masses or uniform in total mass and mass ratio) does not greatly affect our result, but that a larger prior boundary weakens the constraints on the area theorem.

As might be expected, an earlier truncation time ($t_{hmeco}$) for the pre-truncation analysis or a later truncation time ($3\text{ms}$) for the post-truncation analysis gives weaker constraints on the area change.

Comparing different waveform models, we find much stronger constraints on the area theorem for GW150914 using NRSur7dq4 or IMRPhenomXPHM than using IMRPhenomPv2. This may be due to the fact that NRSur7dq4 and IMRPhenomXPHM models include sub-dominant modes, whereas IMRPhenomPv2 does not.

\begin{figure*}[htp]
	\centering
	\begin{minipage}{\columnwidth}
		\centering
		\includegraphics[width=\textwidth]{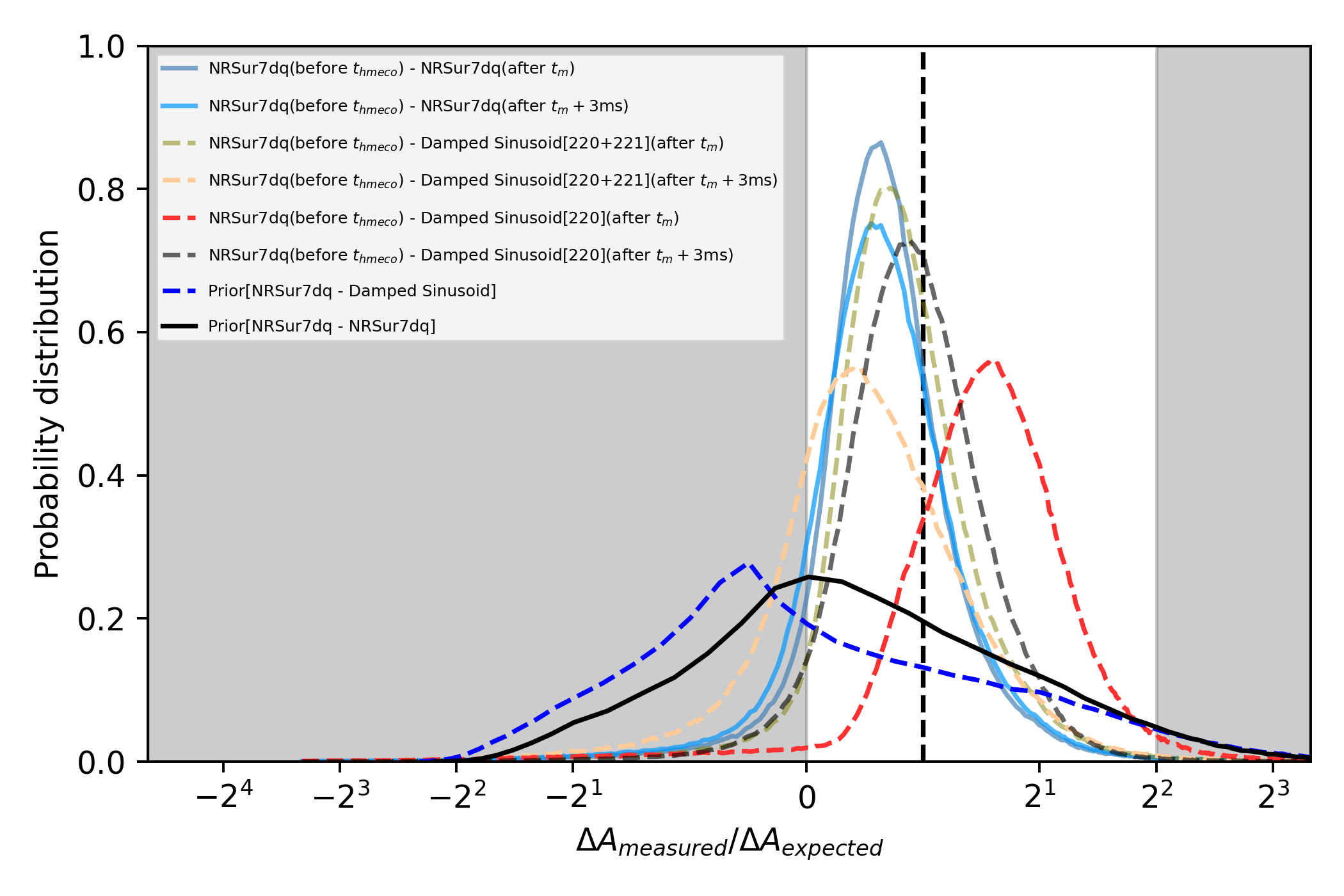}
		\label{Fig:GW150914_area}
	\end{minipage}%
	\hspace{0.3cm}
	\begin{minipage}{\columnwidth}
		\centering
		\includegraphics[width=\textwidth]{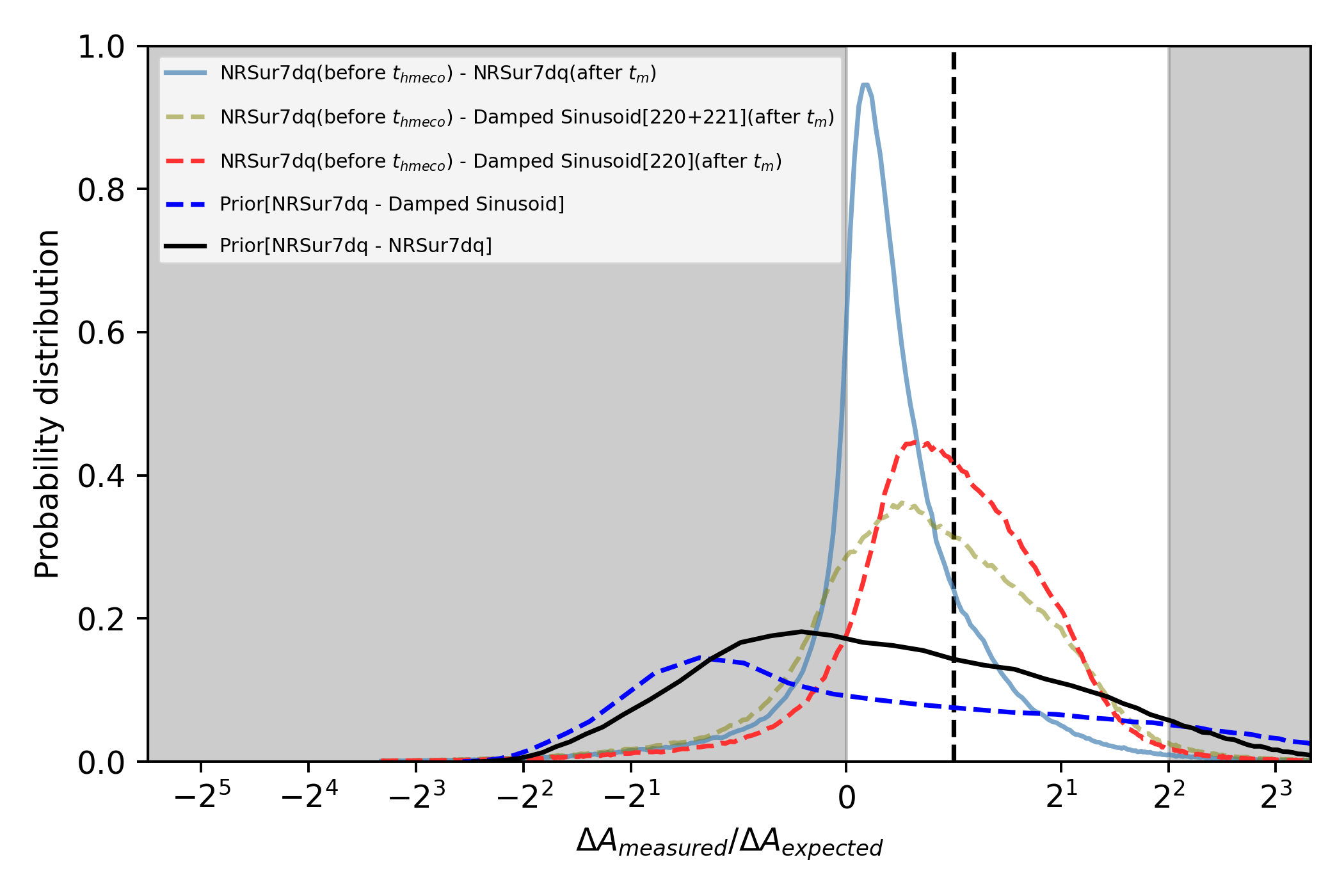}
		\label{Fig:GW170814_area}
	\end{minipage}
	\caption{Ratio of the measured and the expected change in black hole horizon area, $\mathcal{R}=(A^{measured}_{f} - A^{measured}_i)/(A^{expected}_{f} - A^{measured}_{i})$, for GW150914 (left) and GW170814 (right). Here, $A^{measured}_i$ is the initial area inferred using the parameters obtained from the pre-truncation analysis. Shown are results using the NRSur7dq waveform model at different ``{\it gate-start-times}'', as indicated in the legend. The measured final area $A^{measured}_f$ is computed from the post-truncation analysis, using either damped sinusoids ([220], [220+221]) or the NRSur7dq waveform model. The expected final area $A_f^{expected}$ is computed by converting the initial parameters from the pre-truncation analysis (used to compute $A^{measured}_i$) into expected final mass and spin via fits to GR waveforms~\cite{Healy_2014, Hofmann_2016, Jim_nez_Forteza_2017}. The vertical black dashed line denotes the ideal scenario when the expected and the measured change of area are the same. We have also plotted the two different prior distributions used by the blue dashed and the black solid lines.}
	\label{Fig:expected_area_change}
\end{figure*}

For comparison, we also provide a similar study on GW170814 data (see Table~\ref{Area-table}). Compared to the GW150914 results, we find weaker bounds on the area theorem for GW170814. This is due to fact that GW170814 has a quieter post-merger signal. When the pre-truncation analysis is extended only until $t_{hmeco}$, we find that IMRPhenomPv2 provides similar constraints to NRSur7dq4. On the other hand, when the pre-truncation analysis is extended until the merger, the IMRPhenomPv2 analysis provides a stronger bound. We also find a negative change in black hole area when we use IMRPhenomPv2 or IMRPhenomXPHM in the post-truncation analyses after $t_m$. Due to lower SNR ($\sim 7$) for the post-truncation data segment, we obtain larger posterior bounds on the final mass, having support from very low masses (as low as $\sim 20M_\odot$). This leads to a negative change in the BH horizon area.

In order to test the consistency of our results with the actual (positive) change of area predicted by GR, we plot the ratio between the measured change and the expected change in area, $\mathcal{R}=(A^{measured}_{f} - A^{measured}_i)/(A^{expected}_{f} - A^{measured}_{i})$, for GW150914 and GW170814 in Fig.~\ref{Fig:expected_area_change}. 
The expected final area, $A_f^{expected}$ is computed by converting the initial parameters (used to compute $A^{measured}_{i}$) from the pre-truncation analyses to the final parameters assuming GR. All the pre-truncation analyses are performed before $t_{hmeco}$ to completely avoid the merger regime. For demonstration purposes here we have only used NRSur7dq4 as the IMR model in all of the pre-truncation analyses.

If the GR prediction is true, we expect $\mathcal{R}$ to be exactly 1 for a perfect measurement. Due to statistical uncertainties, the probability distribution on $\mathcal{R}$ is expected to be a Gaussian with mean equal to one. The priors on masses and spins used for the different pre- and post-truncation analyses are equivalent to a prior distribution on $\mathcal{R}$ that can be both negative as well as a positive, with $\mathcal{R}>4$ corresponding to a violation of the conservation of energy (see the black solid and the blue dashed lines in Fig.~\ref{Fig:expected_area_change}).

For GW150914, the post-truncation analysis using the [220] damped sinusoid mode starting at $t_m$ produces a probability distribution on $\mathcal{R}$ peaking at a value greater than 1. Consistently, for this particular case we find a better agreement with the area theorem ($\sim 99\%$ [see Table~\ref{Area-table}]).
This is expected, as the area theorem only requires the final area to be larger than the initial area, so the agreement is improved when the distribution for final area is shifted to higher values.
The final mass and spin are overestimated for this case as seen in the right panel of Fig.~\ref{Fig:GW150914_IMR}, resulting in overestimating the final area, with the bias visible in Fig.~\ref{Fig:expected_area_change}.

On the other hand, when we consider the first overtone with the dominant mode after $t_m+3$ ms, we find a shift in the distribution to the opposite direction and the agreement with the area theorem drops. To quantify the agreement with GR, we also quote the mismatch, $C_\mathcal{R}$ for each of the curves in Figs.~\ref{Fig:expected_area_change} in Table~\ref{Area_change-table}. Here the mismatch $C_\mathcal{R}$ denotes the probability of getting $R$ within the range of values symmetric about the median value of $R$, extending to R=1 for each of the curves. To be precise, $C_\mathcal{R}$ denotes the area under the curve bounded by $\mathcal{R}=1$ and symmetric around the mean value.
Lower values of $C_\mathcal{R}$ refer to higher accuracy in recovery of the GR estimates. We find that excluding the merger and starting the post-truncation analysis with [220] at $t_m+3$ms provides a better agreement with GR having $C_\mathcal{R}\sim6.5\%$ as compared to other cases where we use different waveform models (IMR model or the damped sinusoids with [220+221]) after $t_m$ or $t_m+3$ms.

A similar study for GW170814 is also given in Fig.~\ref{Fig:expected_area_change} and Table~\ref{Area_change-table}. As opposed to the GW150914 result, here we find that the dominant mode post-truncation analysis recovers the GR value with greater accuracy, with the mismatch only $C_{\mathcal{R}}=2.9\%$. As opposed to this, using the NRSur7dq model for the post-truncation analysis leads to lower agreement with GR.

\begin{table*}
\begin{center}
\begin{tabular}{|| l | c ||}
 \hline
 \textbf{GW150914} &\shortstack{ $C_\mathcal{R}$\\ (NRSur7dq4)}\\ [0.5ex] 
 \hline\hline
 IMR model before $t_{hmeco}$-Damped Sinusoid [220] after $t_m$ & 67.8\% \\
 \hline
 IMR model before $t_{hmeco}$-Damped Sinusoid [220+221] after $t_m$ & 27.6\% \\
 \hline
 IMR model before $t_{hmeco}$-Damped Sinusoid[220]  after $t_m+$3 ms & 6.5\% \\
 \hline
 IMR model before $t_{hmeco}$-Damped Sinusoid[220+221]  after $t_m+$3 ms & 42.3\% \\
 \hline
 IMR model before $t_{hmeco}$-IMR model after $t_{m}$& 47.5\% \\
 \hline
 IMR model before $t_{hmeco}$-IMR model after $t_{m}+3$ms & 44.6\% \\
 \hline
 & \\
 \hline
 \hline
 \textbf{GW170814} &\\ [0.5ex] 
 \hline\hline
 IMR model before $t_{hmeco}$-IMR model after $t_{m}$ & 61.2\% \\
 \hline
 IMR model before $t_{hmeco}$-Damped Sinusoid[220]  after $t_m$ & 2.7\% \\
 \hline
 IMR model before $t_{hmeco}$-Damped Sinusoid[220+221]  after $t_m$ & 7.9\% \\
 \hline
 \hline
\end{tabular}
\caption{Mismatch $C_\mathcal{R}$ for the recovery of $\mathcal{R}=$1 using the various pre- and post-truncation analyses for GW150914 and GW170814. Lower value of the mismatch $C_\mathcal{R}$, denotes greater agreement with GR predictions. The IMR model used here is NRSur7dq.}
 \label{Area_change-table} 
\end{center}
\end{table*}

\section{Conclusion}
\label{sec7}
In this paper we provide an extensive study on the validity of Hawking's area theorem and the consistency of different phases of the compact binary signal using GW150914 and GW170814 data. We investigate how the different waveform models, various prior distributions, and use of different data segments affect the final results. We observe that uniform priors on component masses or the total mass and mass ratio yield similar results.

For both GW150914 and GW170814, different waveform models provide different constraints on both the final mass and final spin parameters, but all of them are consistent with the full IMR analysis except one. The ringdown analysis of GW150914 at $t_m$ considering only the dominant mode overestimates the final mass and final spin for the remnant black hole. However, considering one additional overtone provides consistent bounds when compared to the result obtained by analysing the full data segment using IMR waveform models. In the case of GW170814, we find that the dominant mode analysis and the [220+221] mode analysis give consistent bounds.

We observe that the different choices for the excision of data and for the waveform models introduce significant systematic errors in the measurements of the validity of the area theorem. In the case of GW150914, for various combinations of excised data, the probability of the validity of the area theorem varies in the range $\sim74\%-94\%$ when we use the NRSur7dq4 waveform model in the pre-truncation analysis. Using instead the waveform model IMRPhenomXPHM, the range drops by $\sim 3\%$, to $\sim 71\%-91\%$. IMRPhenomPv2 leads to an even broader range from $54\%$ to $85\%$.

In Ref.~\cite{Cabero_2018}, a study of a similar test of the area theorem was presented based on a simulated GW150914-like signal. This study used IMRPhenomPv2 for the inspiral and a [220] damped sinusoid for the ringdown, finding support for the validity of the theorem of $\sim 74\%$. Our analysis of the GW150914 data shows a similar result of $\sim 70\%$ when using the corresponding waveform model choices (see 9th row on the 4th column of Table.~\ref{Area-table}).

When considering different waveform models along with the different choices of data duration, the probability on the validity of the area theorem varies in the range $54\%-94\%$. We also see that using a damped sinusoid as the waveform model for the post-truncation analysis improves the probability by $\sim 10\%$ on average as compared to one of the updated IMR waveform models. We also see that if we apply the ringdown analysis with the dominant mode [220] at $t_m$ for GW150914, we get a lower ringdown frequency. This overestimates the final mass, making an area theorem test biased in a direction that likely overstates the agreement, with the probability being $\sim 99\%$. For this reason, we believe the ratio of the measured change in area to the expected change $\mathcal{R}$, and the associated mismatch $C_{\mathcal{R}}$, to be a better metric for determining the consistency of the signal with GR. This yields weaker agreement with GR for the [220] ringdown analysis at $t_m$, for which $C_{\mathcal{R}}\sim 67\%$. This is expected, since a ringdown with only the [220] is not thought to be good model of the signal at merger. For comparison, the [220+221] ringdown analysis at $t_m+3$ms yields $C_{\mathcal{R}}\sim 6.7\%$. 

For GW170814, various data durations used in the analyses lead to larger uncertainty on the probability values as compared to GW150914. For NRSur7dq4 it ranges in between $71\%-87\%$. However, for IMRPhenomXPHM and IMRPhenomPv2 it varies from $17\%-85\%$ and $27\%-90\%$, respectively. To compare between a damped sinusoid and an IMR waveform model for the post-truncation analysis, we find that the damped sinusoid improves the probability of a positive area increase by $\sim 15\%$ as compared to the case where the IMR model is NRSur7dq4. But the improvement could be as high as $\sim 60\%$ when the IMR models used are IMRPhenomXPHM or IMRPhenomPv2. The similar findings are reflected in Table~\ref{Area_change-table} through the values of $C_{\mathcal{R}}$. In this case the best case scenario is obtained using an IMR model before $t_{hmeco}$ and the [220] mode analysis after $t_m$, for which $C_\mathcal{R}= 2.7\%$. However, using IMR model to analyze the post-truncation segment after $t_m$, $C_{\mathcal{R}}$ drops to $\sim 61\%$.

These large systematic uncertainties highlight the need for binary black hole observations across a longer time period than is possible with current generation detectors. This should become possible in the 2030s with the launch of the space-based Laser Interferometer Space Antenna (LISA)~\cite{LISA2017} and the beginning of ground-based ``3G'' detectors,  such as Cosmic Explorer (CE)~\cite{Reitze:2019iox} and Einstein Telescope (ET)~\cite{Punturo_2010}. With a sensitive frequency band of $0.1-100$mHz, LISA will be able to detect binary black holes $\sim$ a year before their merger are detected by ground-based detectors~\cite{Sesana:2016ljz,Barausse_2016,Toubiana:2020vtf,Liu:2020nwz, Carson_2019, Gnocchi_2019, Gupta_2020, Datta:2020vcj}. This should yield unprecedented precision measurements of the fundamental laws governing black hole thermodynamics.

\section{Acknowledgements}

 We thank Bruce Allen,  Alexander H. Nitz and Jahed Abedi for interesting discussions and their valuable inputs. We also thank the Atlas Computational Cluster team at the Albert Einstein Institute in Hanover for assistance. MC acknowledges funding from the Natural Sciences and Engineering Research Council of Canada (NSERC). This research uses the data obtained from the Gravitational Wave Open Science Center (https://www.gw-openscience.org/ ), a service of LIGO Laboratory, the LIGO Scientific Collaboration and the Virgo Collaboration.  LIGO Laboratory and Advanced LIGO are funded by the United States National Science Foundation (NSF) who also gratefully acknowledge the Science and Technology Facilities Council (STFC) of the United Kingdom, the Max-Planck-Society (MPS), and the State of Niedersachsen/Germany for support of the construction of Advanced LIGO and construction and operation of the GEO600 detector.  Additional support for Advanced LIGO was provided by the Australian Research Council.  Virgo is funded, through the European Gravitational Observatory (EGO), by the French Centre National de Recherche Scientifique (CNRS),the  Italian  Istituto  Nazionale  di  Fisica  Nucleare  (INFN)  and  the  Dutch  Nikhef,  with  contributions  by  institutions  from  Belgium,  Germany,  Greece,  Hungary,  Ireland,  Japan,  Monaco,Poland, Portugal, Spain.

\appendix
\setlength{\columnsep}{22pt}
\bibliographystyle{apsrev4-1}
\bibliography{area_theorem}

\end{document}